# Development of trust based access control models using fuzzy logic in cloud computing

Abhishek Kesarwani *, Pabitra Mohan Khilar

*Department of Computer Science and Engineering, National Institute of Technology, Rourkela, India*

ABSTRACT

Cloud computing is the technology that provides different types of services as a useful resource on the Internet. Resource trust value will help the cloud users to select the services of a cloud provider for processing and storing their essential information. Also, service providers can give access to users based on trust value to secure cloud resources from malicious users. In this paper, trust models are proposed, which comes under the subjective trust model based on the behavior of user and service provider to calculate the trust values. The trust is fuzzy, which motivated us to apply fuzzy logic for calculating the trust values of the cloud users and service providers in the cloud environment. We use a Mamdani fuzzy method with gauss membership function for fuzzification and triangular membership function for defuzzification. Parameters such as performance and elasticity are taken for trust evaluation of the resource. The attributes for calculating performance are workload and response time. And for calculating elasticity, we have taken scalability, availability, security, and usability. The fuzzy C-means clustering is applied to parameters for evaluating the trust value of users such as bad requests, bogus requests, unauthorized requests, and total requests.

© 2019 The Authors. Production and hosting by Elsevier B.V. on behalf of King Saud University. This is an open access article under the CC BY-NC-ND license (http://creativecommons.org/licenses/by-nc-nd/4.0/).

## 1. Introduction

Cloud computing (CC) is the technology that provides a facility to store, process, and manage data on remote servers rather than a personal computer or local server. Cloud computing avails different types of resources to the user in the form of service. The NIST (National Institute of Standards & Technology) defines cloud computing as a model for enabling convenient, ubiquitous, on-demand network access to a shared pool of configurable computing resources that can be rapidly released and provisioned with minimal management effort or service provider interaction (Mell et al., 2011). There are mainly three service delivery models, i.e., infrastructure as a service (IaaS), platform as a service (PaaS), and software as a service (SaaS) and four deployment models, i.e., public cloud, private cloud, community cloud, and hybrid cloud available in the cloud environment. Nowadays, security is a big concern in the cloud due to the huge increment in users. There are different techniques to provide security in the cloud, e.g., encryption technique using cryptography, data integrity technique, trust based access control technique, etc. Although not much attention has been given towards the access control model based on the behavior of the system. Under the circumstances that user's and provider's behavior will be changing over time, so trust values also vary accordingly. Our proposed models are based on behavior, which is a good approach towards achieving dynamic access control by calculating trust values from time to time (Behera and Khilar, 2017). Trust based access control provides security from unauthorized access and various security threats, e.g., multi-tenancy cloud platform, DoS, insecure interfaces & API, malicious attacker, abuse of cloud services, data loss, etc. The traditional approach (Behera and Khilar, 2017) has been failed to improve the security requirements of cloud computing. Thus, various access control models have been proposed to improve the traditionally based access control models. So that the dynamic requirements of cloud security can be fulfilled. Traditional access control can only guarantee access to the users in the cloud, but they fail to detect the operation performed by the user, i.e., user behavior. On the other hand, choosing a trusted service provider is also a big challenge. So cloud security has become an important topic among the researchers, and they were always trying to find the different methods to secure the cloud. Some of the researchers mainly focused on the trustworthiness of cloud service providers, and they cover some of the parameters for selecting trusted cloud service providers (Cayirci et al., 2016; Tang and Liu, 2015; Hullermeier and Rifqi, 2009). Whereas our model covers many parameters as possible. In this paper, we have shown how behavioral parameters of different user and service provider are taken into consideration and applied fuzzy methods to calculate trust value and classification of trust. The proposed model is developed for a platform as a service (Paas), which is in a public cloud. There are five types of trust models, i.e., based on agreement, feedback, domain, certificates/keys, and subject. Our proposed model is an improvement of the traditional machine learning model (Khilar et al., 2019), and we also design it for a cloud service provider. The fuzzy based trust model comes under the subjective based model.

\* Corresponding author.
E-mail address: keabhi20@gmail.com (A. Kesarwani).

This paper contains five sections, i.e., first section contain the introduction of our paper, in the second section – we have discussed various existing access models, in the third section – we will present our proposed TBAC model for cloud users and cloud service providers using fuzzy methodology in cloud computing, in forth section – we have focused on the implementation part and experiment results. In this chapter, we also have compared our models with the existing models. We have implemented the proposed model for the diverse number of users and also for CSPs, and in the fifth section – we have discussed the future work and the conclusion part.

## 2. Literature survey

### 2.1. Access control model

Access control is the collection of programs that are used to restrict the user's access. An access control system that records and monitors all the try made by users to access the cloud. Access control also finds out unauthorized access. The access control system is designed with the help of algorithms, models, and administrative capabilities by which every access control system has its own methods, attributes, and capabilities to restrict the user's access (Sun et al., 2011). The main aim of designing this model in a cloud platform is to secure the user's computation and data. Access control decides the type of operation which can be performed by the users on a specific resource and which user has the right to access the resource. In a cloud environment, access control system takes different actions such as authentication, identification before actual accessing of the resource. The basic five access control model which can be applied to the cloud environment is MAC (Mandatory Access Control), DAC (Discretionary Access Control), RBAC (Role Based Access Control), and ABAC (Attribute Based Access Control) are the traditional approach is based on the identity of users while ABE (Attribute Based Encryption) schemes are modified access models which give the concept of encryption. There are also many different type of access models such as Context-Aware Access Control which is the extension part of role-based access control model which is used to manage sensitive data and find whether user's request to limit data access permissions based on the contextual conditions (Trnka and Cerny, 2016; Schefer-Wenzl and Strembeck, 2013; Hosseinzadeh et al., 2016; Colombo and Ferrari, 2017; Kayes et al., 2018; Kayes et al., 2019; Kayes et al., 2017). Trnka and Cerny (2016) have proposed a CAAC scheme based on using security levels, in which the RBAC policies are used to grant and manage data access decision. Schefer-Wenzl and Strembeck (2013) have proposed an ontology-based CAAC approach. Hosseinzadeh et al. (2016) have used OWL language and ontological techniques for modeling context-aware RBAC policies. Colombo and Ferrari have proposed a fine-grained CAAC framework (Colombo and Ferrari, 2017) – designed the access control mechanism for MongoDB, enhancing the data protection functionalities of NoSQL data store. Recently, Kayes et al. have introduced several CAAC models in the last few years (Kayes et al., 2018; Kayes et al., 2019) – that can be applicable in today's IoT-enabled smart spaces and cloud computing environments – for managing the private/sensitive data. They have considered a wide variety of contextual conditions, for example, the situational and relationship context, utilizing the process of inferring implicit knowledge from the currently available context information. A CAAC model for handling imprecise contexts using fuzzy logic and ontology-based approach (Kayes et al., 2017).

Papadakis-Vlachopapadopoulos et al. (2019) discuss the complex networks such as fog computing and mobile edge computing which requires collaborative service level agreement (SLA) and reputation-based trust for a cloud environment. Srivastava and Daniel (2019) design an efficient model for selecting the cloud services on the basis of trustworthiness; they also use fuzzy logic for calculating the cloud trust value.

### 2.2. Trust based access control models

Most of the Internet applications are using cloud technology by using its services such as IaaS, SaaS, and PaaS. Billions of users are using the cloud, and they show very dynamic and uncertain behaviors, then the probability for the existence of malicious behavior is more, so the security of the cloud is really a big concern. The models we have discussed so far are not suitable for a cloud environment. So the researchers developed a new way to access the cloud by using the trust of the users. Every trust based access model works on user behavioral parameters. Various trust models have been proposed, and every trust model used different parameters to evaluate the trust value of a user. Trust based access control not only evaluates for the trust value of users but also for cloud service providers.

#### 2.2.1. Agreement-based trust models

The formation of trust value in this model is on the basis of agreements which are signed by the providers for transferring the various kind of services to cloud users. The trust model is accountable for making and exchanging the agreement on the basis of user specifications. The user specifies different types of QoS (Quality of service) requirements and security to the trust module. The agreement may be in the form of a service or SLA (service level agreement) practice statement (Alhamad et al., 2010). Afterward, the trust module transfers the agreement cooperation request to the service provider as per the user's specified parameters.

#### 2.2.2. Certificate-based trust models

The formation of trust value in this model is on the basis of endorsement keys, trust tickets (TTs), and certificates that are issued by the certificate authority. The most practical and feasible solution to evaluate the trust for infrastructure, platform, and software services are provided by the security certificates. The confidentiality and integrity of data are ensured by the trust tickets (Krautheim et al., 2010). Trust tickets also raise the confidence of the user. To make sure that the control over the user whose data are sent to the cloud computing environment, this model provides a different type of certificates and secret keys. For configuring the cloud and evaluating the measurement of trust, an endorsement key in the trust platform module is used.

#### 2.2.3. Feedback-based trust models

The formation of trust value in this model is done by gathering the opinions and feedback of the users to find whether the service provider is trusted or not. With the help of a service registry module, different service providers are enrolled with the model. Later, the feedback is gathered and managed by the feedback module in which the feedback is given by users about different security parameters and the Quality of Service of the enrolled service providers. On the basis of feedback, the trust value of a service provider is calculated by the model (Habib et al., 2011). The cloud users can see the trust value of a specific service provider by sending the request to the model for a required provider.

### 2.2.4. Domain-based trust models

The formation of trust value in this model is done by partitioning the cloud into two independent domains, i.e., inter-domain trust relationship and within domain trust relationship, which is taken from recommended and direct trust table, respectively (Jamshidi et al., 2013). Mainly in grid computation, this model is used, which is limited in number. If entities are in the same domain and the trust values depend on the transaction happen between these entities, then it comes under within-domain (Li and Ping, 2009). When some entity wants to check the trust value of any entity which may present outside the domain, then it initially searches on a direct trust table, and if the entity is not available on this table, then it checks for recommended trust value which is given by other entities. The trust values provided by the inter-domain are on the basis of recommended trust from different domains and from the direct trust table.

### 2.2.5. Subjective trust models

The formation of trust value in this model is done by dividing the trust into different subgroups, i.e., trust based on code, execution, and authority. The main techniques involved in this model are the fuzzy set and probability set theory for finding the information based on the trust of a specific user or service provider. For calculating the trust values, this model uses either fuzzy or probabilistic theory with given weights (Li et al., 2012). The final trust value of the service providers or users is the combination of the value, which is calculated for each subgroup.

## 3. Proposed access control models

Cloud computing provides a distributed environment, and services are provided to the users based on their use. The role of cloud computing is to give facility to store, process, and manage data on remote servers rather than the personal computer or local server. By this, it prevents the user from installing the costly application on their local machine. And for the demanded services, users do not need to preserve their physical infrastructure. Cloud services are provided by various service providers. The services are in the form of IaaS, SaaS, and PaaS.

It is hard to find whether the user's data which was outsourced for capitation having integrity or not because the user does not have control over their data and its calculation. There might be a chance that an attacker can edit or delete their data or calculation. In order to protect data and computation, CSPs (cloud service providers) need to protect the remote server from various types of threats. The cloud providers and users must be trusted before users send their data and calculation. The cloud service providers trust users on the basis of their status code i.e., 400 (bad request), 401 (unauthorized request), and 403 or 404 (bogus requests). And users can trust upon CSPs on the basis of their elasticity, performance, time, cost, and security. If the behavior of the user is malicious, then they are treated as an untrusted user.

So, for the use of a remote server, there must be a trusted user. If some valid user does some attack i.e., bad behavior, then service providers get affected and not able to fulfill the service level agreements (SLA) and security goal. To secure resources, cloud service providers grant only trusted users. To find whether a user is trusted or untrusted by finding trust values. The user is treated as trusted if the value is greater than the threshold.

### 3.1. Problem description

This paper gives light on the issues related to security, user authorization, and quality of service given by CSPs (Mahallat, 2015). We focus on calculating the trust value of CSPs and cloud users on the basis of cloud resources ability and types of a request by users, respectively. The improvement to a mathematical formula by using a fuzzy approach including fuzzy c-means clustering for finding the trust value of n cloud users set and fuzzy approach based on performance and elasticity to find the trust value of CSPs is given below.

### 3.2. Proposed model

In this paper, we proposed two access control model i.e., cloud user based and cloud service provider based, as shown in Fig. 1 by the use of fuzzy logic, which contains several steps. The proposed model is divided into three parts i.e., user part, service entity part, and management part.

**User Part**

Step 1: User send request for accessing the cloud or to know the trust value of cloud service providers.

Step 2: Behavior monitoring component will analyze whether the request is bad, bogus, or unauthorized.

Step 3: The request of user is stored on the user behavior database.

Step 4: User trust evaluation component apply the proposed technique and finds the trust value of cloud user.

Step 5: The trust value is stored in user trust database.

**Service entity part**

Step 1: Various service entity mentioned their specifications about their resources like ram speed, processor clock speed, etc.

Step 2: SLA supervision component analyze the availability, scalability, response time, and workload of a cloud.

Step 3: Feedback component contains the details of security and usability of cloud given by the cloud users who access it.

Step 4: Monitored SLA database stores all the parameters of each service entity.

Step 5: Elasticity evaluation component finds the elasticity of a cloud using fuzzy logic.

Step 6: Performance evaluation component also use for finding the performance of cloud using fuzzy logic.

Step 7: Resource trust evaluation components use the above two parameters which is elasticity and performance to find out the trust value of CSPs through proposed techniques.

Step 8: CSPs trust value is stored in service entity trust database.

**Trust Management Module (TMM)**

The trust management module is one of the important modules which helps the service entity to know whether the user is trusted or not. And cloud users can also know the trust value of CSPs by sending a request to TMM. It also decides the access to the user either its granted or denied.

An architecture for applying fuzzy logic on the cloud platform is shown in Fig. 2 can be discussed as the technique to provide crisp data from the dataset as an output. The architecture contains the inference rules, fuzzification module, defuzzification module, and decision component. The crisp data are given as input to the fuzzy inference, and these data are converted to a fuzzy logic set using the membership functions of fuzzy. The process of converting is known as fuzzification. Fuzzy rules are used (as it is a rule based model) after fuzzification to get the fuzzy value as an outcome. Afterward, the defuzzification method is applied to change the fuzzy output to the crisp output value. Crisp sets are used most of the time in our life. In the crisp set, an element can be either a member of the set or not. On the other hand, fuzzy sets allow the element to be partially available in the set.

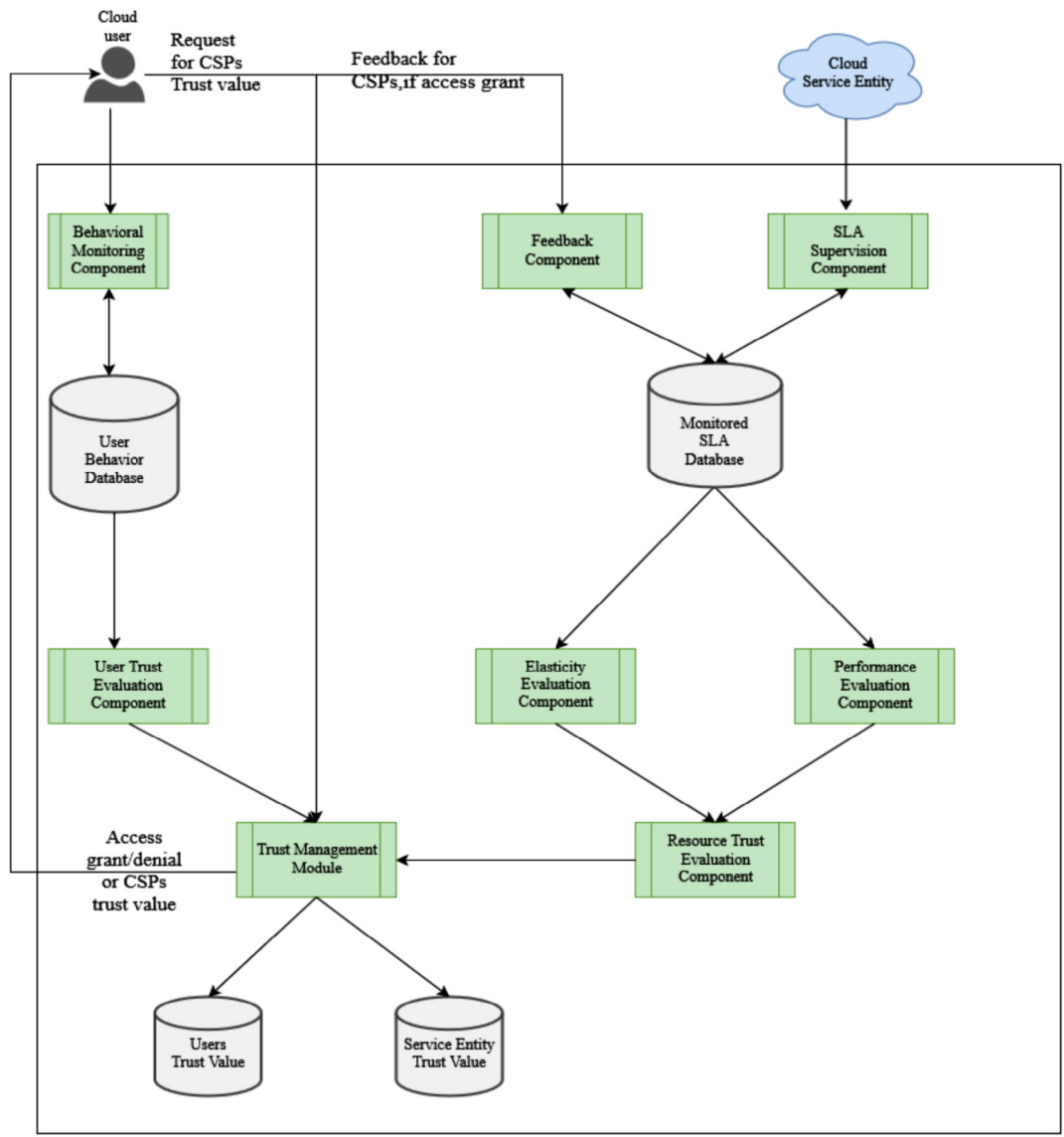

Fig. 1. Flow of requests in Proposed Model.

### 3.3. Preliminaries

We considered two parameters, i.e., performance and elasticity, by which we can find out the trustworthiness of resources. For the performance, we have taken two attributes, i.e., workload and response time. In the cloud environment, the workload is the amount of processing done by the computer at a time. The workload contains the number of application programming running in the cloud and usually the number of users connected to and interacting with the software or application. Response time is the time that was taken by the system to respond to a service that is requested (Kayes et al., 2015). For elasticity, we have taken four parameters, i.e., security, scalability, availability, and usability. Scalability (Ran, 2003) is the ability of a cloud to stay to function well when it is changed in volume or size in order to avail the user's need. Availability (Manuel, 2015), in respect of a cloud, denotes the ability of users to access the information or resources in the right format and in a specified location. Usability is the degree of ease with the cloud, which can be used to achieve the required goals efficiently and effectively. Usability assesses the problems involved in using a user interface. Security in cloud com-puting is termed as the collection of policies designed to follow the rules and protect the infrastructure associated information, and data application with cloud use. For the user side, we use four parameters, i.e., Bad requests (400), Bogus requests (404), Unauthorized requests (401 and 403), and total requests. HTTP status codes 400 in which the server will not process the request due to a spacious client error (e.g., a malformed syntax of the request, framing invalid request message, too large size of the request or deceptive request routing). Bad requests are the types of the request which could not be known by the servers due to malformed syntax. The purpose of these requests is to consume the bandwidth for denial of service attacks. 401 similar to forbidden it occurs when authentication is required and has not yet been provided or has failed. Unauthorized requests mean that users are trying to access those resource for which he/she has not proper authorization. It indicates that users are trying stealing or modifying data, and 404 occurs when requested resources could not be found but may be available in the future.

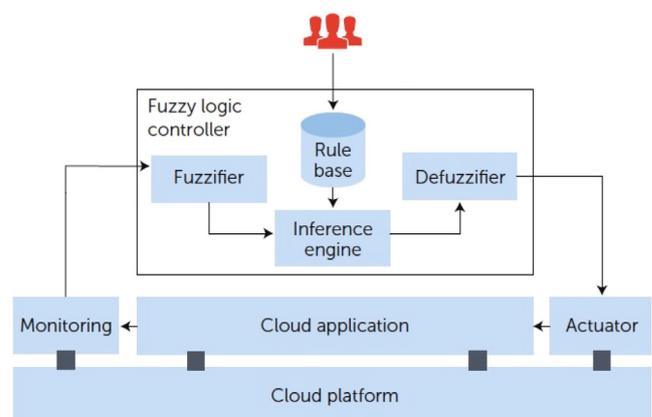

Fig. 2. Fuzzy logic architecture in Cloud Computing.

### 3.4. Cloud trustworthiness for service user

This section proposes a method to compute the trust value for cloud users. Trust value of the cloud users will show user's behavior in the cloud. In this method, to evaluate the trust value of users, several parameters have been taken into considerations. We take the average of bad requests, bogus requests, and unauthorized requests. The weight factor is given to all average requests to set the effectiveness of that specific request. Then we subtract the total negative value by one to get the trust value.

To ease the description, we first define some notations:

- **UAR**: denotes the number of "Unauthorized requests" performed by the user over a period T.
- **BOR**: denotes the number of "Bogus requests" performed by the user over a period T.
- **BAR**: denotes the number of "Bad requests" performed by the user over a period T.
- **TR**: denotes the number of "Total requests" performed by user over a time T.
- **T_neg**: denotes the negative trust value.
- **UT**: denotes the user trust value.

The unauthorized request means that users are trying to access those resources for which he has not proper authorization. It indicates that users are trying stealing or modifying data. The unauthorized request rate (UARR) is calculated via Eq. (1).

$$UARR = \frac{UAR}{TR} \quad (1)$$

Bogus requests are the request used by the attacker to waste the processor cycle. The main purpose of this request is to waste the CPU cycle of the remote system. Bogus request rate (BORR) is calculated via Eq. (2).

$$BORR = \frac{BOR}{TR} \quad (2)$$

Bad requests are the types of the request which could not be understood by the servers due to malformed syntax. The main purpose of these requests is to consume the bandwidth for denial of service attacks. So this parameter is used to check the availability of the cloud. Bad request rate (BARR) is calculated via Eq. (3).

$$BARR = \frac{BAR}{TR} \quad (3)$$

In this, we find the trust value evaluation formula is given below in Eq. (4 and 5):

$$T\_neg = w1 \times UARR + w2 \times BORR + w3 \times BARR \quad (4)$$

$$UT = 1 - T\_neg \quad (5)$$

where w1, w2 and w3 are weight factors such that w1 + w2 + w3 = 1 given by Behera and Khilar (2017) as w1 = 0.5, w2 = 0.2 and w3 = 0.3. We apply the same weight values as taken by Behera and Khilar (2017) for comparison with our FIS model.

The fuzzy logic theory is used so as to increase the mathematics ontology in a certain method with fuzziness in order to make an intelligent decision. We apply fuzzy c-means clustering (Nayak et al., 2015) on total requests, bad requests, bogus requests, unauthorized requests with respect to trust value to get the fuzzy sets and divide all requests into 25 clusters. These are the linguistic variable.

#### 3.4.1. Fuzzy inference system

In Matlab, the fuzzy logic toolbox is used to design and develop our model, which contains the evaluation methods to implement many types of fuzzy inference such as the Sugeno and Mamdani inference system. In the proposed model, we use a Mamdani fuzzy method with gauss membership function (Jain et al., 2016) of range 0 to 1 for inputs the clusters to convert it into the fuzzy set from the crisp set as fuzzy Inference system is rule based so we have to provide rules for evaluating the outputs, i.e., trust value as shown in Table 1. The rules are combined together to get the fuzzy output. Then, we convert the fuzzy output into crisp output by the defuzzification process. We use a triangular membership function that refers to appropriate fuzzy weights, which lie between 0 and 1 (Nagarajan et al., 2018).

### 3.5. Cloud Trustworthiness for service provider

This section proposes a method to compute the trust value for cloud service providers. The trust value of the cloud service providers will show service entity behavior in the cloud. In this method, to evaluate the trust value of service providers, several parameters have been taken into considerations, i.e., availability, scalability, usability, security, workload, and response time. For our proposed model, we use FIS editor in Matlab with two input factors: performance and elasticity. These two inputs are directed as crisp inputs to the fuzzy inference. The fuzzy logic methodology is used for performance having two inputs, i.e., workload and response time. The five fuzzy sets, which are very low, low, medium, high, and very high, are used to characterize the fuzzy value for workload, and for response time, there are also five fuzzy sets, i.e, instantaneous,

**Table 1**
Samples of fuzzy rules for users trust evaluation.

| IF Bad requests | and Unauthorized requests | and Bogus requests | and Total request | Then Trust |
|---|---|---|---|---|
| Cluster 1 | Cluster 1 | Cluster 1 | Cluster 1 | Cluster 1 |
| Cluster 2 | Cluster 2 | Cluster 2 | Cluster 2 | Cluster 2 |
| – | – | – | – | – |
| – | – | – | – | – |
| – | – | – | – | – |
| Cluster 25 | Cluster 25 | Cluster 25 | Cluster 25 | Cluster 25 |

fast, medium, slow, and very slow. The fuzzy sets that present the output parameters are: high, medium, and low. And for elasticity calculation, we have four inputs: security, scalability, availability, and usability. These four parameters are given as input to the fuzzy inference system with membership function, as shown in Fig. 4. The fuzzy logic method in this report uses three fuzzy sets for the inputs and five fuzzy sets for the parameters of output. These three fuzzy sets, which are high, medium, and low are used to characterize the value of fuzzy for every input factors which are, security (S), usability (U), availability (A), and scalability (Sc). These fuzzy sets present the output parameters as excellent, very good, good, very poor, and poor. Table 2 shows samples fuzzy rules for input factors and the assigned values for elasticity (E) evaluation. Fig. 5 shows the membership function of elasticity. There will be 81 rules for the evaluation of elasticity. Table 3 and 4 show the response time and workload quantification, respectively, for applying the rule to calculate the performance. And all 25 combinations of fuzzy rules for performance evaluation are shown in Table 5. Fig. 6 shows the membership function of performance. Then, we combine the result of elasticity and performance to evaluate the trustworthiness of the service provider. Table 6 shows the sample rules of a fuzzy set for calculating trust value. There will be 15 rules for calculating the trustworthiness of CSPs. We divide the trust into three parts, i.e., low, medium, and high. The user's feedback is required for calculating the trust value of a cloud service provider,

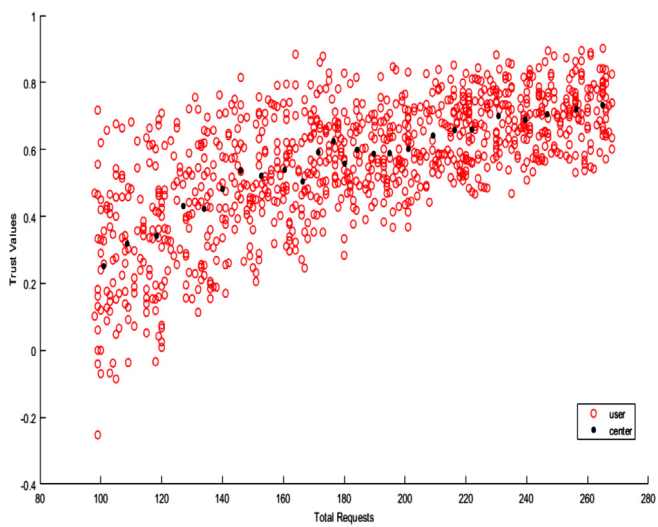

**Fig. 3.** Clustering form between the trust value and total requests.

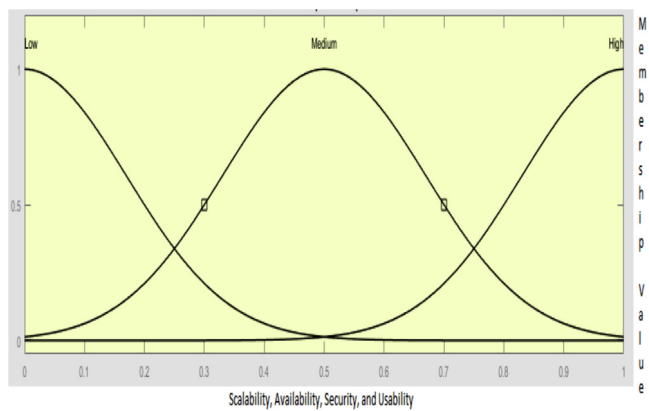

**Fig. 4.** Membership function for scalability, availability, security, and usability.

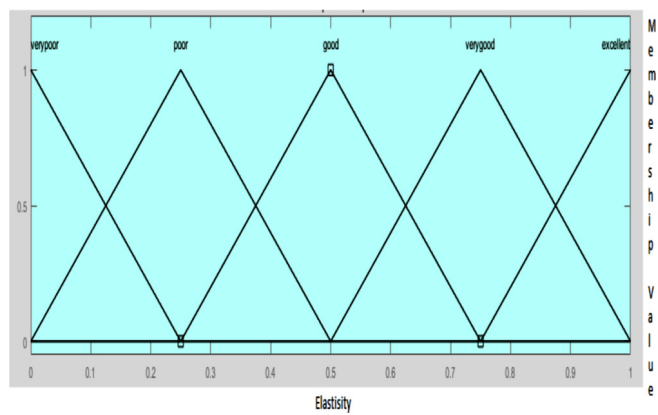

**Fig. 5.** Membership function for elasticity.

**Table 3**
Workload quantification.

| Workload (per process) | Mean | | Standard Deviation | |
|---|---|---|---|---|
| | Start | End | Start | End |
| Very Low | 0 | 26 | 0 | 8.2 |
| Low | 23 | 41 | 7.2 | 6.95 |
| Medium | 37 | 65 | 5.8 | 3.9 |
| High | 62 | 83 | 4.6 | 6.7 |
| Very High | 79 | 100 | 6.3 | 0 |

**Table 2**
Fuzzy rules sample for elasticity evaluation.

| IF Sc | and A | and S | and U | Then E |
|---|---|---|---|---|
| Low | Low | Low | Low | Very Poor |
| Medium | Low | Low | Medium | Poor |
| Medium | Medium | Low | Medium | Good |
| Low | Medium | Medium | Low | Poor |
| Medium | Low | Medium | Medium | Good |
| High | Low | Low | High | Poor |
| Medium | Medium | Medium | Medium | Good |
| High | High | Low | High | Good |
| Low | High | High | Low | Good |
| High | Low | High | High | Very Good |
| High | Medium | Medium | High | Good |
| Medium | High | Medium | Medium | Good |
| High | Medium | High | High | Very Good |
| High | High | High | High | Excellent |

**Table 4**
Response time quantification.

| Response time (in milliseconds) | Mean | | Standard Deviation | |
|---|---|---|---|---|
| | Start | End | Start | End |
| Instantaneous | 0 | 7.1 | 0 | 5.2 |
| Fast | 6 | 19 | 4.1 | 5.3 |
| Medium | 18.5 | 40.5 | 5.5 | 8.5 |
| Slow | 37.5 | 62.5 | 7.1 | 9.4 |
| Very Slow | 60 | 100 | 7.8 | 0 |

**Table 5**
All combination of Fuzzy rules for Performance evaluation.

| IF Workload | and Response Time | Then Performance |
|---|---|---|
| Very low | Very slow | Low |
| Very low | Slow | Low |
| Very low | Medium | Medium |
| Very low | Fast | Medium |
| Very low | Instantaneous | Medium |
| Low | Very slow | Low |
| Low | Slow | Medium |
| Low | Medium | Medium |
| Low | Fast | Medium |
| Low | Instantaneous | Medium |
| Medium | Very slow | Medium |
| Medium | Slow | Medium |
| Medium | Medium | Medium |
| Medium | Fast | Medium |
| Medium | Instantaneous | High |
| High | Very slow | Medium |
| High | Slow | Medium |
| High | Medium | Medium |
| High | Fast | High |
| High | Instantaneous | High |
| Very high | Very slow | Medium |
| Very high | slow | Medium |
| Very high | Medium | High |
| Very high | Fast | High |
| Very high | Instantaneous | High |

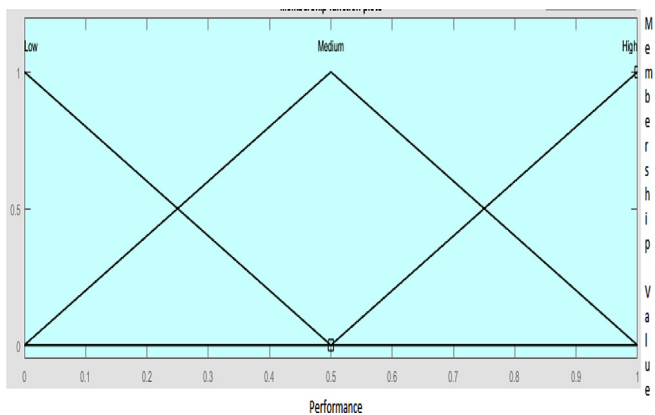

Fig. 6. Membership function for performance.

**Table 6**
Fuzzy rules sample for trust evaluation of CSPs.

| IF Performance | and Elasticity | Then Trust |
|---|---|---|
| Low | Very Poor | Low |
| Low | Good | Low |
| Low | Excellent | Medium |
| Medium | Poor | Low |
| Medium | Good | Medium |
| Medium | Very Good | High |
| High | Very Poor | Medium |
| High | Good | High |
| High | Excellent | High |

if we get negative feedback greater than 40%, then we ban this service provider. We choose 40% as a higher threshold because some users may give fake feedback.

## 4. Simulation results and experimental setups

In this, a trust evaluation result based on the fuzzy logic system is shown. The proposed scheme enables to evaluate the trust values for cloud users and service providers. In this section, we have shown the clustering of users, surface view diagrams for user side trust model, and providers side trust model. By using 1000 users data, we apply fuzzy c-means clustering with $c = 25$ with different features. The clustering between the trust value and total requests shown in Fig. 3.

Fig. 7–9 shows the surface view of bad requests, bogus requests, and unauthorized requests with respect to total requests respectively. And Fig. 10–13 shows the surface view of usability and scalability, security and availability, workload and response time, performance and elasticity respectively. The proposed fuzzy model

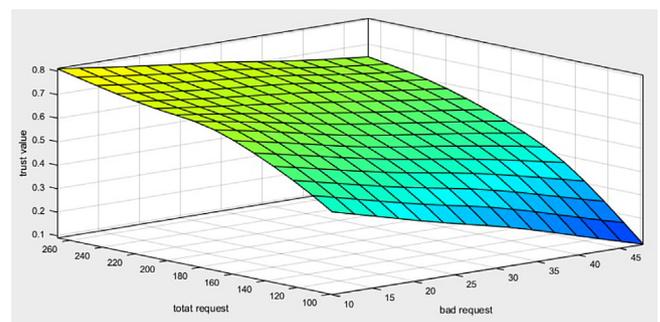

Fig. 7. Surface view between bad requests and total requests.

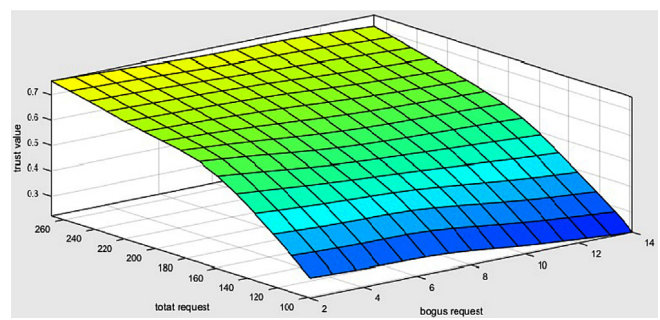

Fig. 8. Surface view between bogus requests and total requests.

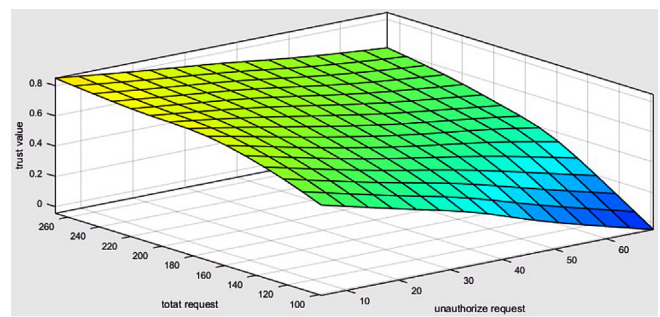

Fig. 9. Surface view between unauthorized requests and total requests.

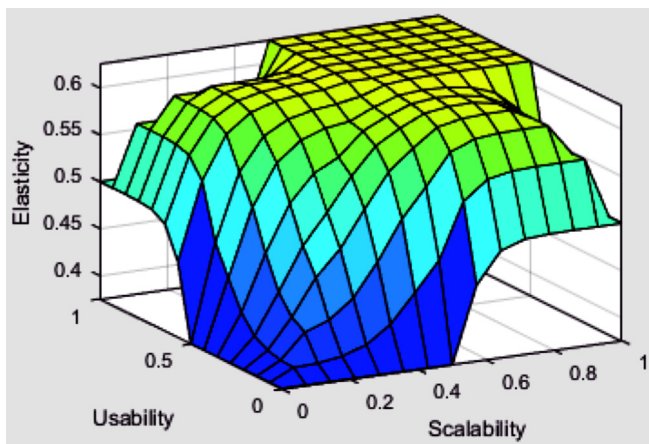

**Fig. 10.** Surface view between usability and scalability.

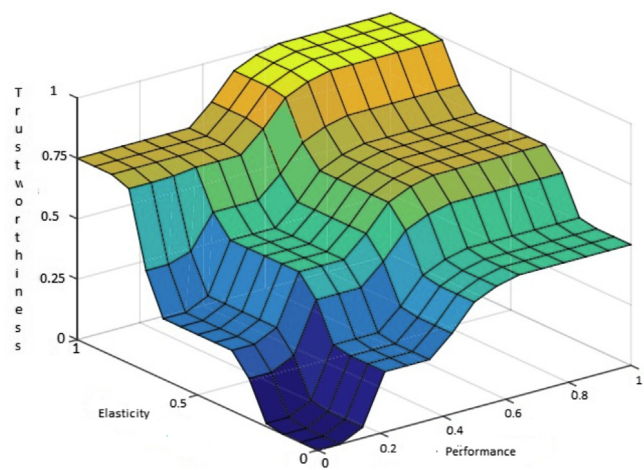

**Fig. 13.** Surface view between performance and elasticity.

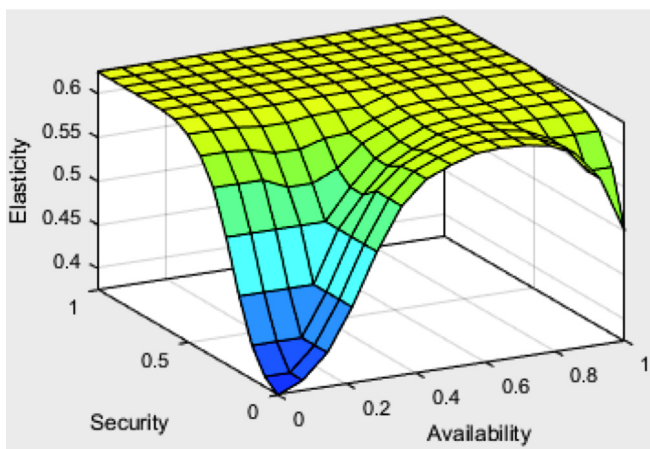

**Fig. 11.** Surface view between security and availability.

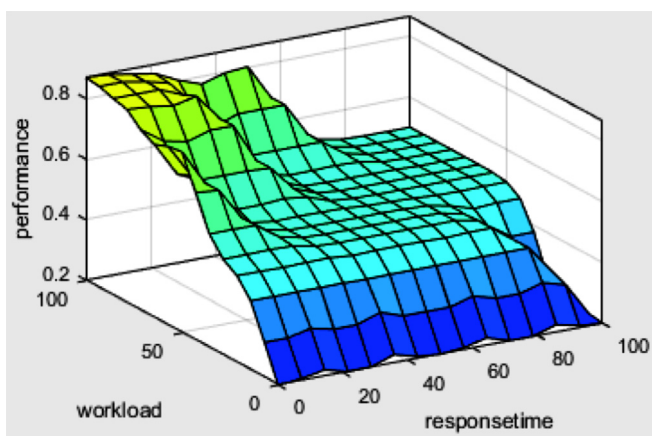

**Fig. 12.** Surface view between workload and response time.

can be applied for the use of additional input parameters. Moreover, the proposed model can also be applicable to different types of web-based application such as online shopping websites, etc.

We use MATLAB software (Mathworks, Natick, MA) for our experiments, and we have used 1000 training data and 300 testing data, which is in CSV format. For cloud user datasets, we have five fields i.e., bad requests, bogus requests, unauthorized requests, total requests, and user trust value as given in Eq. (5). For the cloud service provider, we use Gaussian distribution on usability, scalability, security, and availability for elasticity. And for performance, we apply Gaussian distribution on workload and response time. We collect this type of data on a local server, and in benchmarking datasets, we get only 1 or 2 malicious users among a large amount of users, so we don't go for it.

All scenarios of cloud user behavioral parameters are plotted with respect to the total requests are shown in Fig. 7–9. For CSPs, we can say that the usability and scalability of CC must be high for the better elasticity of the cloud, as shown in Fig. 10. And availability has some threshold value because if a cloud environment accepts more jobs and not able to complete these jobs will affect the elasticity of CC. The security of the CC is directly proportional to the elasticity, as shown in Fig. 11. For better performance of CC, the workload must be high, and response time must be low, as shown in Fig. 12. Then, we find the trustworthiness of CSPs with the help of elasticity and performance, as shown in Fig. 13. Finally, we have compared the user based model with the previous mathematical model in Fig. 15 and our FIS model, also give much better result as compare to different existing techniques as given in Table 7.

### 4.1. Comparison

We apply our proposed fuzzy model for 300 test data sets and compare it with the trust, which is evaluated by the mathematical model, and we find that 223 are trusted users, and 77 are untrusted users. The root mean square error (rmse) is 0.0251. Fig. 14 shows trust value for four users and Fig. 15 shows the comparison between both trust values of 300 test data. With the help of the proposed model, we get the trust value of good users is higher than the mathematical model, and for malicious users, it evaluates the trust value, which is lower as compared to the value evaluated by the mathematical model. This is due to the clustering of parameters. By this, we conclude that the proposed model gives a more accurate value than the mathematical model. Table 7 shows the comparison of our model with different traditional machine learning models (Khilar et al., 2019). The time taken by our model is 0.12 s. The mean absolute error and root mean square error of our model are 0.134 and 0.2512, respectively.

Table 7
Comparison with existing techniques.

| Algorithms | Time(s) | MAE (%) | RMSE (%) | Precision | Recall | F1-Score |
|---|---|---|---|---|---|---|
| KNN | 0.12 | 28.92 | 58.15 | 0.75 | 0.74 | 0.74 |
| Nearest-Centroid | 0.26 | 54.7 | 89.91 | 0.62 | 0.58 | 0.58 |
| Gaussian NB | 0.009 | 42.85 | 77.48 | 0.71 | 0.66 | 0.65 |
| Decision Tree | 0.109 | 34.7 | 65.4 | 0.70 | 0.69 | 0.69 |
| Linear SVC | 1.859 | 49.07 | 71.62 | 0.45 | 0.52 | 0.40 |
| Logistic Regression | 0.385 | 39.89 | 69.31 | 0.65 | 0.64 | 0.63 |
| Ridge Classifier | 0.250 | 40.08 | 67.57 | 0.65 | 0.63 | 0.61 |
| FIS (Proposed) | 0.12 | 13.4 | 25.12 | 0.69 | 0.71 | 0.70 |

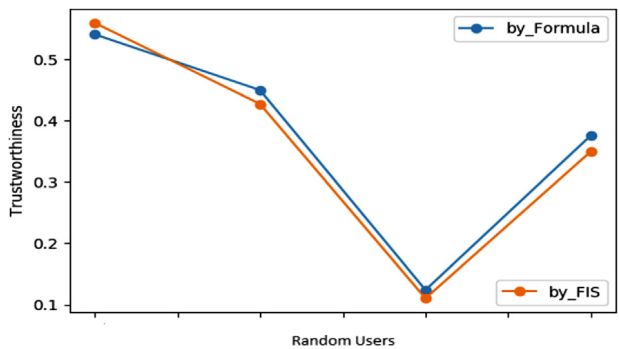

**Fig. 14.** Comparison between trust values of 4 users.

Table 8
Attributes for different parameters.

| Parameters | Attributes |
|---|---|
| Elasticity | Flexibility, Capacity, Portability, Adaptability, Extensibility |
| Performance | Accuracy, Functionality, Interoperability, Stability |
| Time | Users response time, Users request serving time |
| Cost | Ongoing cost, Profit or cost sharing, Acquisition and training cost |
| Data Security | Data recovery, Data location |

the basis of trust value. We have divided the users into two categories, i.e., good and bad users. We have calculated the trust value of CSPs on the basis of QoS parameters and user feedback. We combine the result of elasticity and performance to evaluate the trust values of the CSPs. For finding elasticity, we have four inputs: security, scalability, availability, and usability. And for performance calculation, we have two inputs: workload and response time. By the trust values of service users, we are controlling the access permission of users for accessing cloud resources. CSPs trust values are used to control access for providing cloud services to service users.

### 5.2. Future work

Our future work is to deploy a mutual access control model based on the trust in which users can get access by the fuzzy values accordingly and in this also considered the past trustworthiness, and the user can take the service provider based on their trust. There are some other parameters as given in Table 8 for service provider's which will be included and necessary to introduce a model in the future for optimization of the rules. Now our proposed model for the cloud service provider only focuses on the performance and elasticity. There are also some models available which work on elasticity (Fox et al., 2009). We want to combine all the parameters, i.e., performance, elasticity, cost, time, and data security for the evaluation of the trust value of each service provider by focusing on its attributes, respectively.

### Declaration of Competing Interest

The authors declare that they have no known competing financial interests or personal relationships that could have appeared to influence the work reported in this paper.

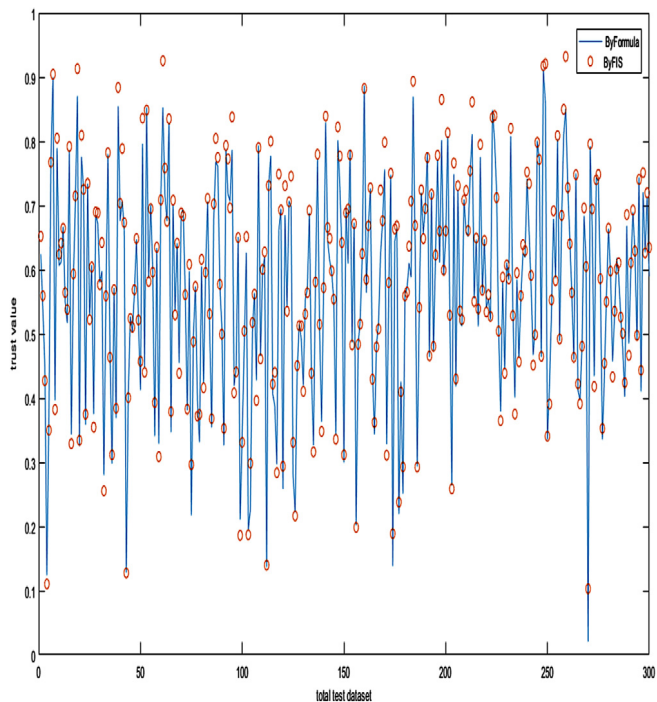

**Fig. 15.** Comparison between trust values of 300 users.

## 5. Conclusion and future work

### 5.1. Conclusion

Trust based access control model is an effective way of securing the cloud environment. In this paper, we have proposed two trust based access control models, i.e., user based and CSPs based using the fuzzy technique. The main purpose of these models is to find the trusted resource for cloud users and authorize the user on